\documentclass[slac_one]{revtex4}
\usepackage{graphicx}
\usepackage{fancyhdr}
\begin{document}

\title{{\small{2005 International Linear Collider Workshop - Stanford,
U.S.A.}}\\ 
\vspace{12pt}
New Physics at the TeV Scale and Precision Electroweak Studies} 

\author{Stephen Godfrey}
\affiliation{Ottawa Carleton Institute for Physics, 
Department of Physics, Carleton University, Ottawa K1S 5B6 Canada}

\begin{abstract}
In this summary we review some recent developments in New 
Physics at the TeV scale.  We concentrate on measurements at the ILC 
that can distinguish between some of the models that have recently 
been discussed, concentrating on results 
presented at this workshop: The 
Little Higgs Model, models of Large Extra Dimensions; Randall-Sundrum 
(RS), Arkani-Hamed Dimopoulos Dvali (ADD), and Universal Extra Dimensions 
(UED).  Some recent results on constraining 
effective Lagrangian parametrizations 
of new physics are also presented.

\end{abstract}

\maketitle

\thispagestyle{fancy}


\section{INTRODUCTION} 

There is a universal consensus that the standard model is a low energy 
effective theory and that some form of new physics exists 
beyond the standard model (BSM).  The literature is full of candidate 
theories but it will be experiment that shows the way.  This 
contribution reviews some of the recent developments in BSM 
phenomenology
with an emphasis on results presented in the BSM working group 
sessions.  However, it would be a mistake to consider this topic
in isolation from the other working group topics.  
I expect the next few years to be exciting times in 
particle physics with the start of the LHC leading to 
major advances in our understanding of electroweak symmetry breaking.  
It is likely that this will give us the first hints of physics BSM but 
it is possible that the first hints will come from elsewhere, perhaps
anomalous properties of the top quark. Maybe this new physics is 
supersymmetry.  And maybe the new physics unravelled at the ILC will 
explain some of the puzzles in cosmology.  The point is that while the 
physics topics have been divided up into EWSB, SUSY, Top/QCD, New 
Physics, and Cosmology, they are all connected and one should not 
forget this when focusing on individual topics.  This will be apparent 
in some of the examples chosen to describe the search for new physics.

There are numerous models of new physics.  An important task of the 
ILC will be to disentangle the possibilities and identify the correct 
one.  We start with a very brief overview of some of the possibilities, 
focusing on models of recent interest.  In the remainder of this 
summary I will examine some of approaches discussed to understand the 
underlying physics.  This summary should not by any means be viewed as a 
comprehensive overview;  it is a snapshot of a selection of topics 
covered at this workshop and studied over the last few years.  More 
detailed reviews are given by LHC/LC working group in 
Ref.~\cite{Weiglein:2004hn}.  See also 
Ref.~\cite{Dawson:2004xz,Moortgat-Pick:2005cw,Lykken:2005up}.
An important omission in this summary is the subject of Higgless 
Elecroweak Symmetry Breaking.  A selection of recent papers on this 
subject is given in Ref.~\cite{Csaki:2003zu}

\section{MODELS OF NEW PHYSICS}

There are many models of new physics.  Some of the models that have 
attracted attention recently are the Little Higgs model
\cite{Arkani-Hamed:2002qy}, 
various models of extra dimensions \cite{Antoniadis:1990ew,Rizzo:2004kr}; 
ADD \cite{Arkani-Hamed:1998nn}, RS \cite{Randall:1999ee}, 
UED \cite{Appelquist:2000nn}, and the Higgsless model \cite{Csaki:2003zu}.  
However,  we shouldn't ignore older models that, 
although less fashionable, may contain elements of truth in them.  
Some models of continued interest are extended gauge sectors with 
extra $U(1)$ factors \cite{Leike:1998wr}
like the $E_6 \to SU(5) \times U(1)_\chi \times 
U(1)_\psi$ where the extra $U(1)$ factors give rise to extra neutral 
gauge bosons, the left-right symmetric model, $SU(2)_L\times 
SU(2)_R\times U(1)$, and dynamical symmetry breaking models such as
technicolour and topcolour \cite{Hill:2002ap}.
From the point of view of disentangling these possibilities we need to 
understand what they have in common and how we can distinguish them.
As a result I will focus on predictions of the various models and how we 
can unravel the underlying physics rather than on theoretical details of 
the various models.  

In the next few paragraphs I will give a rather superficial survey 
of some recent models and refer the interested reader to the 
literature.   My purpose here is to simply identify the 
characteristics of the various models that might reveal themselves by 
experiment.

\begin{description}
\item[Little Higgs Models]  \cite{Arkani-Hamed:2002qy}
are a new approach to 
stabilize the weak scale against radiative corrections. They predict 
new gauge bosons $W_H^\pm$, $Z_H$, and $B_H$ 
and a new heavy top quark at the TeV scale. The parameters of the 
Littlest Higgs model relevant to the discussion are 
$f$, the vev that breaks the global 
symmetry group to a smaller group and sets the mass scale of the new 
heavy particles in the model, and gauge boson mixing angles
$s$ and $s'$.  A light Higgs boson is expected at ${\cal O}(100)$~GeV.  
The couplings to $\gamma\gamma$ are sensitive to new physics running 
in the loop so measurement of the Higgs $\gamma\gamma$  and $gg$ BR's
is a sensitive test of the heavy top quark, extra gauge bosons and new 
scalar particles expected in Little Higgs models.  Other modifications 
are expected due to mixing of TeV-scale particles with SM particles 
and corrections to SM parameters.

\item[Extra Dimensions] \cite{Rizzo:2004kr}
In most scenarios our 3-dimensional space is a 
3-brane embedded in a $D$-dimensional spacetime.  The basic signal is 
a Kaluza Klein (KK) tower of states corresponding to a particle 
propagating in the higher dimensional space-time.  The details depend 
on the geometry of the extra dimensions with many different variations.
\begin{itemize}
\item
The ADD scenario \cite{Arkani-Hamed:1998nn} has a KK tower of graviton 
states in 4 dimensions that behaves like a continuous spectrum which 
Hewett \cite{Hewett:1998sn} parametrized as the effective operator
\begin{equation}
i {{4\lambda} \over{M_H^4}} T^{\mu\nu}T_{\mu\nu}
\label{eq:hewett}
\end{equation}
that will lead to deviations in $e^+e^-\to f\bar{f}$ dependent on 
$\lambda$ and $M_H$, a cut-off scale on the summation over the KK 
states.
ADD also predicts graviscalars and gravitensors propagating in extra 
dimensions.  The parameters of interest for ADD are the mixing 
between the Higgs boson and the graviscalar, $\xi$, the number of 
extra dimensions, $\delta$, and the $M_D$ scale. 
The Mixing of the graviscalar with the Higgs boson leads 
to a significant invisible width of the Higgs.

\item 
In the Randall Sundrum Model \cite{Randall:1999ee}
2 3+1 dimensional branes are 
separated by a 5th dimension.  It predicts the existence of a radion 
which corresponds to fluctuations in the size of the extra dimension. 
Radions have anomalous couplings with gluon and photon pairs and since 
they can mix with the Higgs boson this alters the  corresponding Higgs
BR's.  In the RS model the KK graviton spectrum is discrete and 
unevenly spaced.  It is described in terms of two parameters, the 
mass of the first KK state and the coupling strength of the graviton.
TeV scale graviton resonances are expected to be
produced in 2-fermion channels.
  
\item
In the Universal Extra Dimension scenario \cite{Appelquist:2000nn}
all SM particles propagate in the bulk resulting in KK towers for 
the SM particles with spin quantum numbers identical to SM particles.  
The resulting spectrum resembles that of SUSY and the 
conservation of KK number at tree level ensures that the lightest KK 
partner is stable and is always pair produced so that the signatures 
look alot like signatures of SUSY.

\item 
An extension to both ADD and RS is the existence of higher curvature 
terms in the action for 
gravity that may manifest themselves as we approach $M_{pl}\sim 1$~TeV
\cite{Rizzo:2005xr}. 
In RS the dominant effect is a modification of the KK 
graviton mass spectrum and their couplings to matter.  
In the case of ADD the modifications are quite different.  The usual 
ADD signatures remain unaltered but the modifications 
lead to the production of long lived black holes.
Both the RS and ADD modifications can be studied at the ILC.

\end{itemize}

\end{description}

To summarize, the models predict extra Higgs bosons (doublets and 
triplets), radions, graviscalars, gravitons, KK excitations of the 
$\gamma$, $Z$, $W$, extra gauge bosons and extra fermions. Almost all 
of the models predict new $s$-channel structures at the TeV scale, 
either as extended gauge bosons or new resonances.  
To sort out the models we first need to elucidate and complete the TeV 
particle spectrum and to then make precision measurements of their 
properties.  

\section{PRECISION ELECTROWEAK MEASUREMENTS}

There are several paths to discovering new physics.  The most 
straight forward is the direct discovery of new particles.  The next 
possibility is the indirect discovery by comparing deviations from the 
SM to specific new models.  The final approach is to test for new 
physics by measuring the parameters of effective Lagrangians.  

A starting point for indirect searches for new physics is to
consider the common features of the 
various models.  In almost all cases a new $s$-channel structure is 
expected at the TeV scale either in the form of extra gauge bosons or 
as some other type of new resonance.  Each of these models predicts
different properties for these new resonances so to 
distinguish between the possibilities we will need to make precision 
measurements.  While it is likely that discoveries at the LHC will get 
us started it is almost a certainty that we will need the ILC to 
discriminate between models.  An incomplete list of 
tools we will have at the ILC are measurements of the various di-fermion 
channels, anomalous fermion couplings, anomalous gauge boson couplings  
and measurement of the Higgs couplings.

Let's start by considering the possibility that the LHC discovers an 
$s$-channel resonance in the dilepton invariant mass distribution.  
There are numerous possibilities of what it 
might be; graviton, KK excitations, a $Z'$, etc.  The LHC can give 
some information about what it might be using the invariant mass 
distribution and forward backward asymmetry measurements
\cite{Rizzo:2003ug,Davoudiasl:2000wi,Dittmar:2003ir}.  However, 
these measurements are rather crude and would require significant 
luminosity to achieve any sort of precision and resolving power.  On 
the other hand if we assume the LHC discovers a single, rather heavy 
resonance, the ILC can make many precision measurements such as cross 
section and widths (depending on the mass for the latter case), 
angular distributions, and its couplings via decays and polarization 
measurements.  

\begin{figure*}[h]
\centering
\rotatebox{90}{\includegraphics[width=57mm]{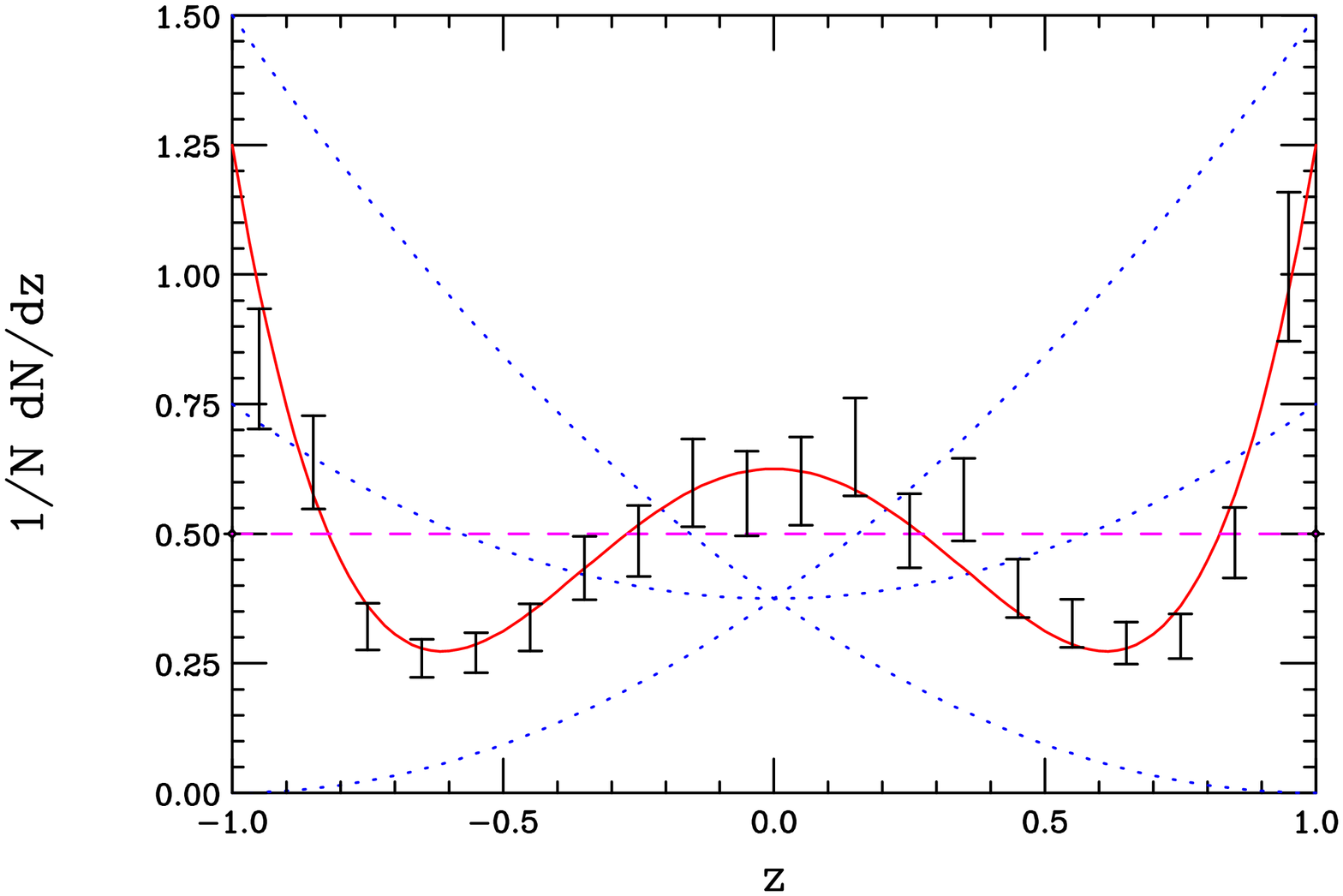}} \qquad
\rotatebox{90}{\includegraphics[width=57mm]{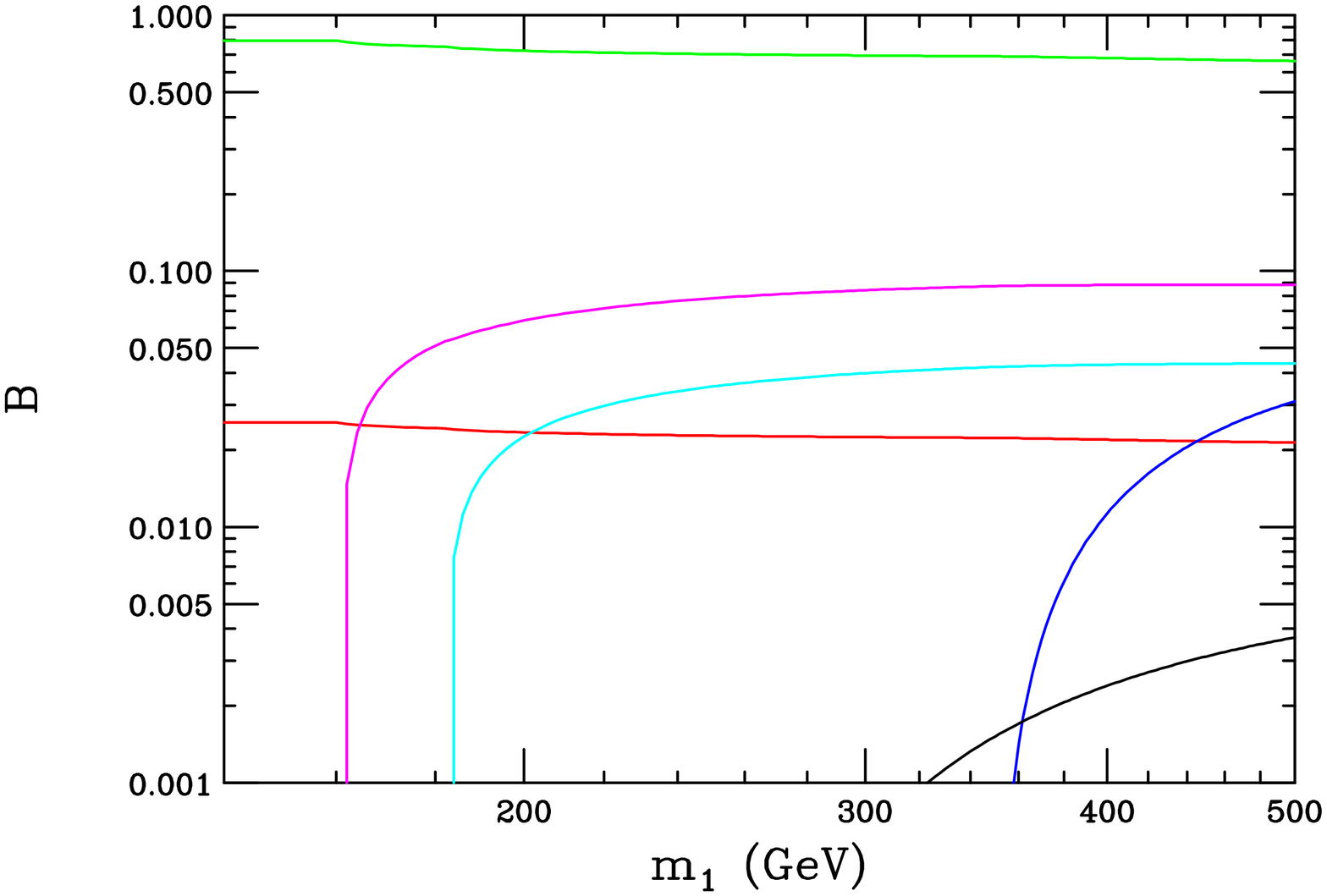}}
\caption{Left side: Normalized angular distribution ($z=\cos\theta$) 
for the decay of a spin-2 graviton 
into fermion pairs (the `w'-shaped curve) in comparison to similar decays by 
either spin-0 (dashed) or spin-1 (dotted) particles. The data with errors show 
the result from a typical sample of 1000 events.
Right side: 
Mass dependencies of the two-body branching fractions for the first 
graviton KK state in the case where the SM fields are on the wall.
From top to bottom on the right side 
of the figure the curves are for dijets, $W$'s, $Z$'s, tops, dileptons and 
Higgs pairs assuming a Higgs mass of 120 GeV.
From Davoudiasl, Hewett and Rizzo \cite{Davoudiasl:2000wi}.} 
\label{davoudiasl}
\end{figure*}

If the resonance is below the ILC threshould it can be produced on 
resonance.  In this case angular distributions can be tested against 
different spin hypothesis to distinguish between a spin 2 graviton and 
say, a spin 1 $Z'$.  BR's could be used to measure the resonance 
couplings which would distinguish between the universal couplings of a 
graviton or the unique couplings expected for the various $Z'$ 
scenarios.  Angular distributions and branching ratios for spin-2 
gravitons from Davoudiasl, Hewett and Rizzo \cite{Davoudiasl:2000wi}
are shown in Fig.~\ref{davoudiasl}.

\subsection{Precision Measurements Using Di-Fermion Channels}

A more likely scenario is that the mass of a new state is 
beyond the direct reach of the ILC.  In this case we can still 
learn a considerable amount about a new resonance.  There are numerous 
di-fermion channels and since the couplings to each 
channel is model dependent, observables such as cross sections to 
specific final states, forward-backward asymmetries, and left-right 
asymmetries can be used to distinguish between models.  

\begin{figure*}[b]
\centering
\includegraphics[width=80mm]{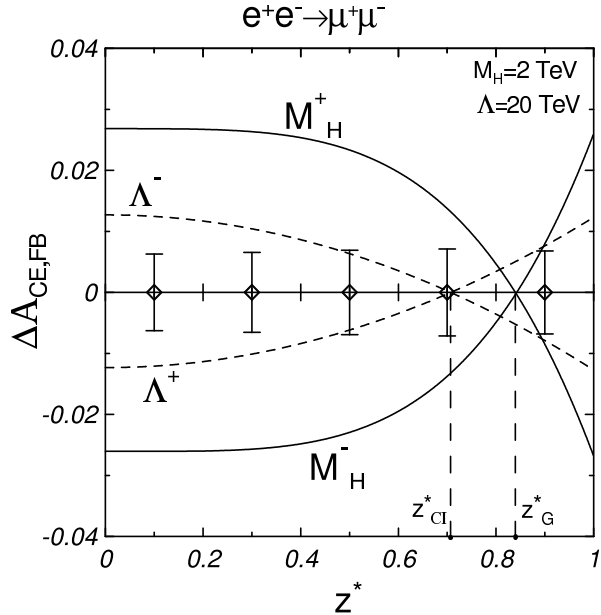}
\caption{$\Delta A_{\rm CE,FB}(z^*)$ in the CI and the ADD scenarios
for the indicated values of $\Lambda$ and $M_H$. The $\pm$ superscripts 
refer to positive and negative interference, respectively. The vertical bars
represent the statistical uncertainty at a LC with $\sqrt s=0.5$ TeV and
${\cal L}_{\rm int}=50\, fb^{-1}$. 
From Osland Pankov and Paver \cite{Pankov:2005qi,Pankov:2005ar}} 
\label{pankov}
\end{figure*}

A first step is to disentangle the spin of the exchange particle.  As  
a concrete example there have been a number of studies examining how 
to distinguish virtual graviton KK expected in the ADD scenario of 
finite size extra dimensions from other possibilities. Hewett 
\cite{Hewett:1998sn}
parametrized the exchange of virtual graviton KK states as 
the effective operator given in Eqn.~\ref{eq:hewett}.
She showed how interference with SM amplitudes leads to deviations in 
the dilepton observables dependent on both $\lambda$ and $M_H$. Rizzo
has studied how multipole moments could be used to distinguish spin 2 
from spin 1 \cite{Rizzo:2002pc}.  
Osland Pankov and Paver used the various difermion 
observables to estimate to what extent $M_H$ could be constrained at 
the ILC \cite{Osland:2005ee,Osland:2003fn}.  
More recently they constructed a 
``Forward-Backward Centre-Edge'' defined as 
$\sigma_{CE,FB}=\sigma_{C,FB}-\sigma_{E,FB}$ to identify graviton 
exchange and act as a discriminator 
between possible models \cite{Pankov:2005qi,Pankov:2005ar}.  
In this defination ``centre'' refers to a 
region with $|\cos\theta| \leq Z^*$ and  ``edge'' refers to 
$|\cos\theta| > Z^*$ with $z^*$ a value that can be varied to optimize 
the discrimination power and $FB$ is the forward-backward asymmetry 
evaluated for the centre region and the edge region. $A_{CEFB}$ is 
shown in Fig. \ref{pankov} for a contact interaction and 
ADD graviton exchange.   They estimated that the ILC would be 
sensitive to $M_H$ up to 3.5 and 5.8~TeV for $\sqrt{s}=0.5$ and 
1.0~TeV respectively with ${\cal L}_{int}=500$~fb$^{-1}$.

\begin{figure*}[t]
\centering
\includegraphics[width=65mm]{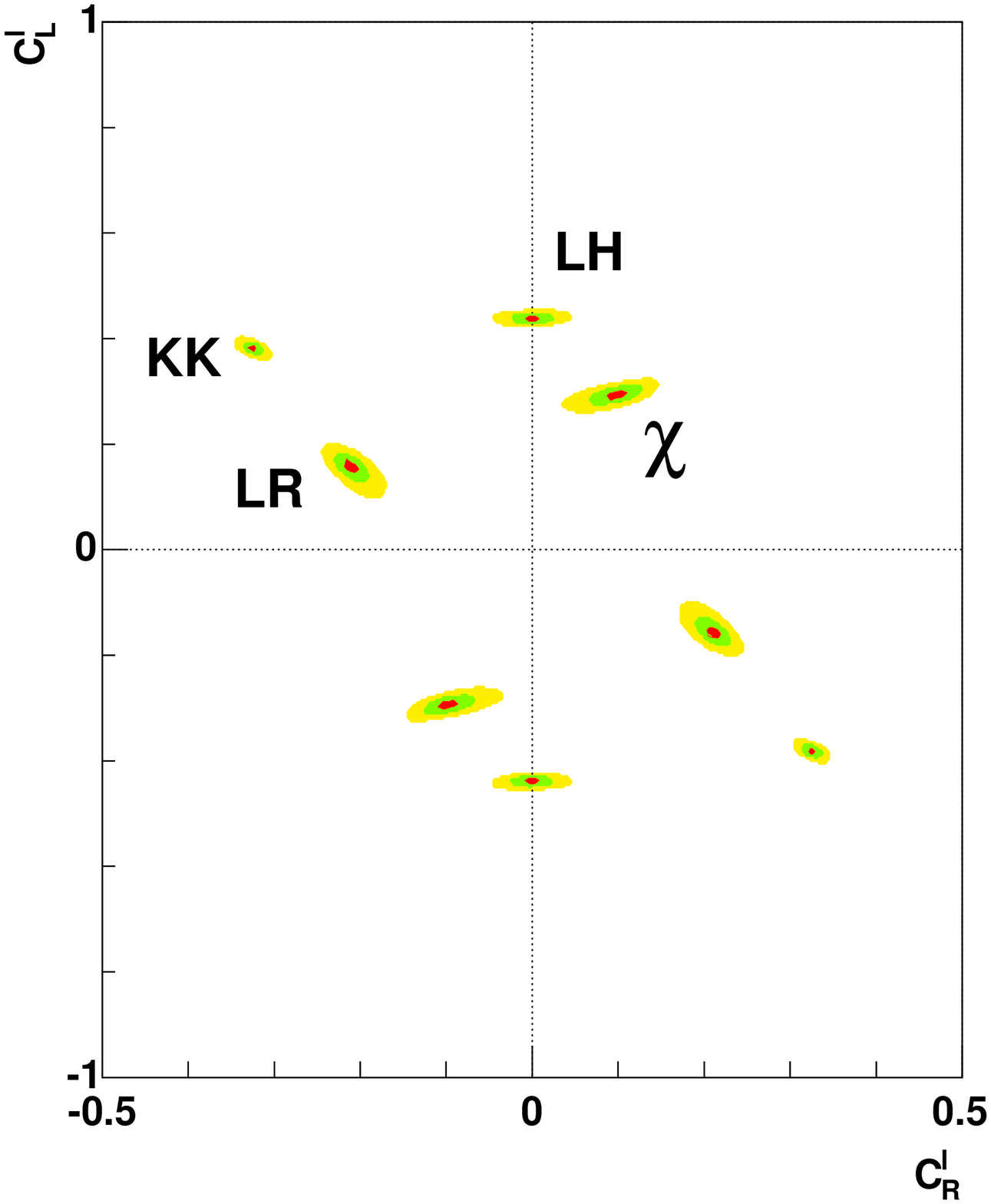}\qquad\qquad
\includegraphics[width=65mm]{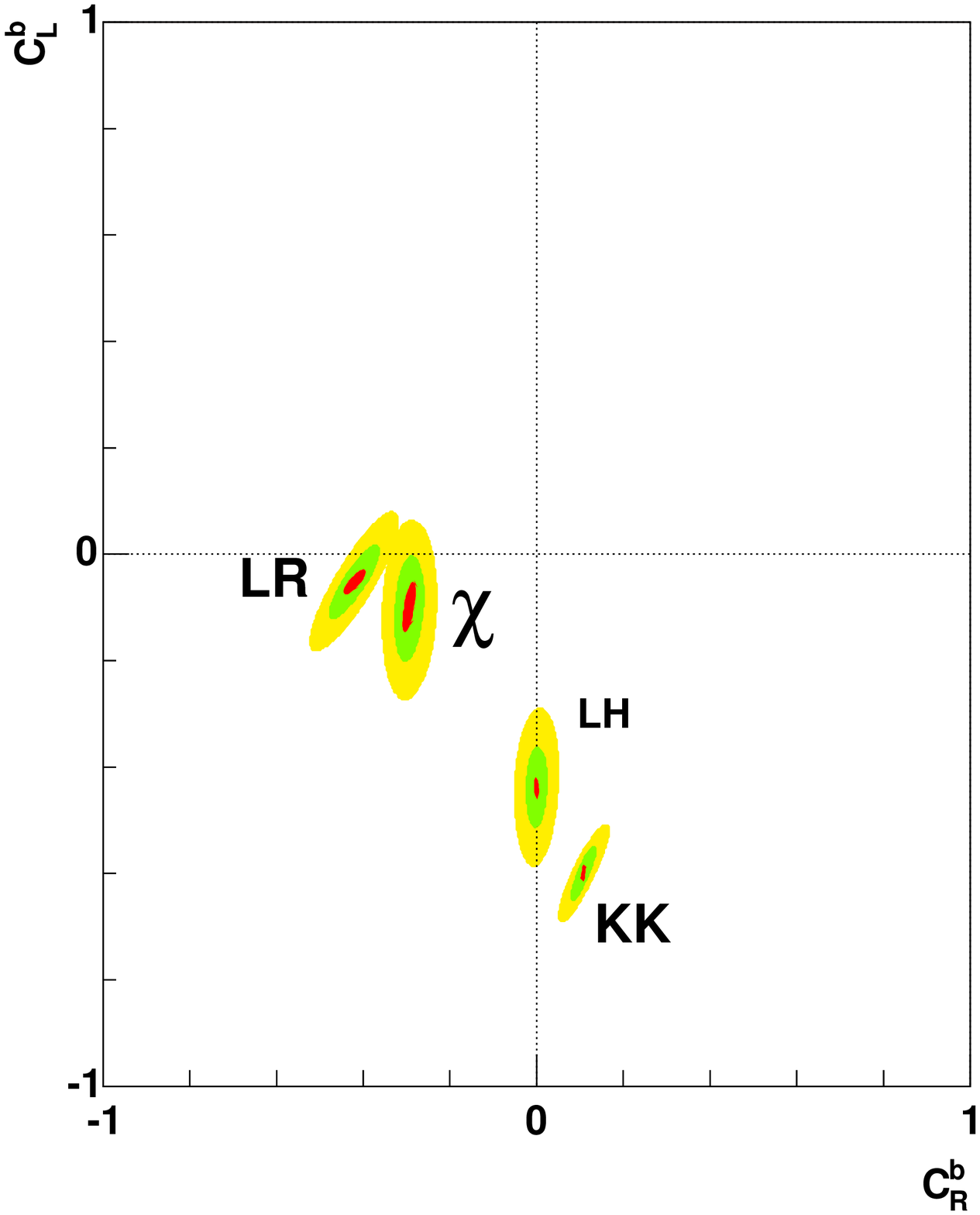}
\caption{Resolving power (95\% CL) for $M_{Z'}=1$, 1.5, and 2~TeV
and $\sqrt{s}=500$~GeV, ${\cal L}_{int}=1$ab$^{-1}$.
The left side is for leptonic couplings based on the leptonic 
observables $\sigma$, $A_{LR}$, $A_{FB}$.  The right side is for $b$ 
couplings based on $b$ observables $\sigma$, $A_{FB}$, $A_{FB}(pol)$
assuming that the leptonic couplings are known and a $b$-tagging 
efficiency of 70\%. The couplings correspond to the $E_6$ $\chi$, LR, 
LH, and KK models.
From Ref. \cite{godfrey}.} \label{godfrey}
\end{figure*}

A next step would be to measure the resonance couplings.  
Riemann used 
the leptonic observables to demonstrate that one can extract a $Z'$ 
couplings and discriminate between models 
\cite{Aguilar-Saavedra:2001rg}.  
A more recent analysis is shown in Fig.~\ref{godfrey} which shows 
the resolution power for $Z'$'s coming from the $E_6$ $\chi$, 
LR-symmetric, Littlest Higgs, and KK excitations. 
Note that the couplings shown for the KK case do not in 
fact correspond to the KK $Z'$ couplings as in this model there are 
both photon and $Z$ KK excitations.  The point is simply the KK model
can be distinguished from other models. 
These figures were 
produced for $\sqrt{s}=500$~GeV and ${\cal L}_{int}=1$ab$^{-1}$
assuming electron and positron polarization of 80\% and 60\% 
respectively, $\Delta P_{e^\pm}=0.5\% $, $\Delta {\cal L}=0.5\%$, and 
$\Delta ^{sys}\epsilon_{lepton}=0.25\%$.  There is a two-fold 
ambiguity in the signs of the lepton couplings since all lepton 
observables are bilinear products of the couplings. The hadronic 
observables can be used to resolve this ambiguity since for this case 
the the quark and lepton couplings enter the interference terms 
linearly.  

\begin{figure*}[b]
\centering
\includegraphics[width=100mm]{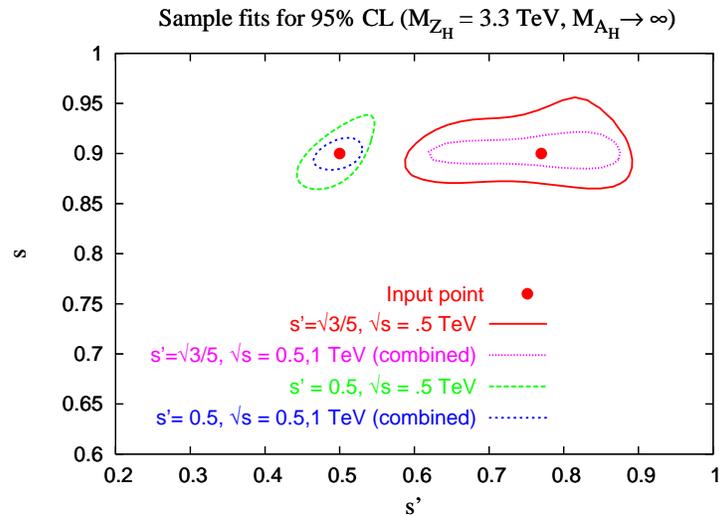}
\caption{95\% CL sample fits to the data points ($s=0.9$, $s'=0.5$)
and ($s=0.9$, $s'=\sqrt{3/5}$) using $e^+e^-\to f\bar{f}$ observables
at a 500~GeV ILC, taking $M_{Z_H}=3.3$~TeV and $M_{A_H}\rightarrow\infty$.
Also shown for each point is an improved fit from adding data from 
a $\sqrt{s}=1$~TeV, ${\cal L} = 500\hbox{fb}^{-1}$ run at the ILC. 
From Conley, Le, and Hewett  \cite{Conley:2005et}.} \label{conley}
\end{figure*}

A complementary approach was described by Conley Le and Hewett 
\cite{Conley:2005et} who showed how the Little Higgs parameter space 
can be probed in the di-fermion channels. Their results are 
shown in Fig.~\ref{conley} which indicates how well the 
parameters of the model can be constrained assuming that the mass of 
the $Z_H$ is known from the LHC.

The di-fermion channel has also been applied to other possible new physics. 
Riemann studied 
the sensitivity to the UED model level 2 KK $Z$ excitations 
in the di-fermion channels \cite{Riemann:2005es}.  
Due to  KK number conservation the KK 
excitations only couple to conventional fermions through loops.  As a 
result the couplings are much smaller than SM couplings and 
$\sigma(e^+e^-\to f\bar{f})$ is much less sensitive to UED KK 
excitations than to other types of new gauge bosons.

\begin{figure*}[t]
\centering
\includegraphics[width=80mm]{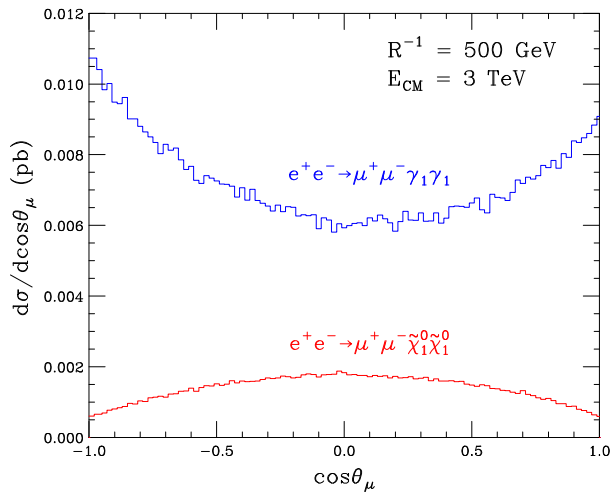}
\caption{Differential cross-section $d\sigma/d\cos\theta_\mu$ 
for UED (blue, top) and supersymmetry (red, bottom)
as a function of the muon scattering angle $\theta_\mu$.
From 
Battaglia, Datta, De Roeck, Kong and Matchev \cite{Battaglia:2005ma}.} 
\label{battaglia}
\end{figure*}

A final example is that of Universal Extra Dimensions (UED).  The KK 
spectrum in UED closely resembles that of SUSY.  The typical signal 
for SUSY is missing $\not{E}_T$.  So, for example, if a signal with 
significant  $\not{E}_T$  was observed at the LHC it is quite possible 
that one could not decide if it was due to SUSY or UED
\cite{Battaglia:2005zf,Battaglia:2005ma}.  
However the spins of SUSY particles are 
different than that of UED KK particles.  One can take advantage of 
this by studying the angular distributions of the outgoing muons in 
$e^+e^-\to \mu^+\mu^- +\not{E}_T$. In UED this signature arises from 
KK muon production,
$e^+e^-\to \mu_1^+\mu_1^- \to \mu^+\mu^- \gamma_1 \gamma_1$ while in 
SUSY it arises from smuon pair production,
$e^+e^-\to \tilde{\mu}^+\tilde{\mu}^- \to \mu^+\mu^- 
\tilde{\chi}_1^0 \tilde{\chi}_1^0$.
For UED the resulting muon angular distribution 
goes like $1+\cos^2\theta$ while for SUSY it goes like 
$1-\cos^2\theta$.  The ISR-corrected theoretical predictions for the 
angular distributions for UED and SUSY are shown in 
Fig.~\ref{battaglia}. Clearly, the two cases can be discriminated. 
In addition to angular distributions threshold 
scans and energy distributions can also be used as discriminators of 
UED and SUSY \cite{Battaglia:2005zf,Battaglia:2005ma}.   In particular, 
the threshold cross section for $e^+e^-\to \mu^+\mu^- +\not{E}_T$
goes like  $\beta^3$ 
$(\beta=\sqrt{1-M^2/E^2_{beam}})$ in the MSSM and like $\beta$ in UED
\cite{Battaglia:2005zf,Battaglia:2005ma}.

As an aside one should not forget the 
considerable experimental effort that goes into these 
measurements.  For example, $b$-tagging is an extremely powerful tool 
for ID'ing models so the understanding $b$-purity versus efficiency is 
an important issue to understand.  Studies of this and other vertex 
detector issues were presented by 
Hillert.  Luminosity and beam parameter measurements was another 
important issue described by  Ingbir and Torrence.

\subsection{Precision Measurement of Higgs Boson Properties}

In addition to measurements of difermion observables, precision 
measurements of Higgs properties can be another important 
discriminator of models.  Higgs properties have been studied in a wide 
variety of models using many different processes. In a first example 
Lillie studied Higgs properties in the Randall Sundrum model (RS)
\cite{Lillie:2005pt}. In 
this model, Higgs production is enhanced at the LHC and in 
$\gamma\gamma$ collisions but reduced at the ILC.  A probably more 
distinctive signal is that Higgs decays are substantially modified 
from their SM values.

Battaglia {\it et al} studied ADD at the ILC \cite{Battaglia:2004js}. 
Mixing of the SM Higgs with the graviscalar induces 
an invisible width compared to direct SM decay.  The ILC can measure this 
invisible width directly and using $HZ$ production.  The invisible 
width can be deduced in the $HZ$ process by observing the $Z$ boson and
reconstructing the missing energy recoiling against it. 
The number of extra dimensions can be constrained by measuring the 
$e^+e^- \to \gamma +\not{E}$ at different values of $\sqrt{s}$ 
\cite{Weiglein:2004hn}. Combined with a Higgs mass measurement from 
the LHC can constrain the ADD parameter space.

\begin{figure*}[t]
\centering
\rotatebox{270}{\includegraphics[width=65mm]{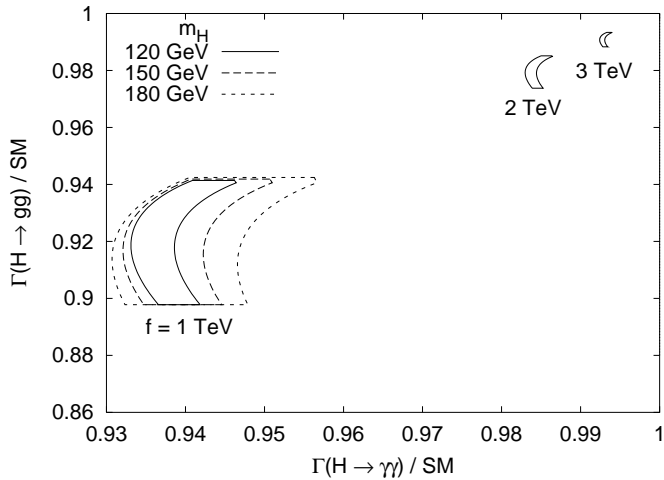}}
\caption{Range of values of 
        $\Gamma(H \to \gamma\gamma)$ accessible in the LH model as a
        function of $f$, normalized to the SM value, for $m_H = 120$,
        150 and 180 GeV and $f=1$, 2, and 3~TeV.
From Han, Logan, McElrath and Wang \cite{Han:2003gf}.} \label{logan}
\end{figure*}

Precision measurements of the Higgs partial widths are a another 
powerful tool for distinguishing models.  The two-photon and two-gluon
partial widths are modified by heavy particles running in the loop and 
by shifts to the SM $W$-boson and $t$-quark masses \cite{Han:2003gf}.  
Fig. \ref{logan} shows the range of $\Gamma (H\to gg)$ versus 
$\Gamma(H\to \gamma\gamma)$ normalized to the SM values for different 
Higgs masses and values of the decay constant parameter of the Little 
Higgs model \cite{Han:2003gf}.  It is clear that precision 
measurements of the BR's offers a good means of constraining the 
parameters of this model. 

\begin{figure*}[b]
\centering
\includegraphics[width=90mm]{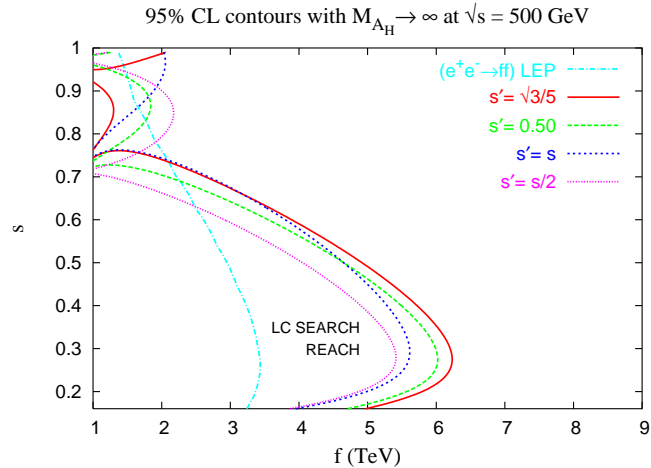}
\caption{The ILC search reach from the process $e^+e^-\rightarrow Z_L h$
    for various values of $s'$, taking $\sqrt{s}=500$ and $M_{A_H}\rightarrow\infty$.
    The LEP II exclusion region from
    $e^+e^-\rightarrow f\bar{f}$ is shown for $s'=s/2$
     for comparison. 
From Conley, Le, and Hewett \cite{Conley:2005et}.} \label{conley2}
\end{figure*}

Conley Le and Hewett also studied the process $e^+e^-\to Zh$ in the 
Little Higgs model \cite{Conley:2005et}.  One of the hallmarks of 
Little Higgs models is the coupling of heavy gauge bosons to $Zh$.  
Thus,  a signature of the Little Higgs model is deviations from the 
SM in $\sigma(e^+e^-\to Zh)$. The sensitivity of this process to the 
parameters of the model are shown in Fig.~\ref{conley2} for 
$\sqrt{s}=500$~GeV and assuming ${\cal L}_{int}=500$~fb$^{-1}$ which 
gives $\delta \sigma_{Zh}/ \sigma_{Zh}=1.5\%$.
The  $\sigma(e^+e^-\to Zh)$ measurement 
covers a large region of the parameter space but there are 
still large regions to explore so perhaps other observables might be 
useful. In any case the 
$\sigma(e^+e^-\to Zh)$ measurements would be a useful complement to 
the difermion measurements in some regions of the LH parameter space 
and would provide a confirmation of this hallmark feature of the LH 
model.

\section{PRECISION MEASUREMENTS AND EFFECTIVE LAGRANGIANS}

In the examples given so far we assumed specific models and examined 
how well precisions measurements could either detect them or 
discriminate them from other models.  A more general approach is to 
use the language of effective Lagrangians 
\cite{Krstonosic:2005qp,Kilian:2005bz,Osland:2005ee}.  
Generically ${\cal L}_{eff}$ can be written:
\begin{equation}
{\cal L}_{eff}={\cal L}_{SM}+\sum_i {{c_i}\over{\Lambda^p} }
{\cal O}^{(4+p)}
\end{equation}
where $\Lambda$ reflects the scale of new physics and the details of the 
new physics (couplings, chiral structure etc.) are parametrized in the 
coefficients $c_i$.  For example, new interactions such as $s$-channel
$Z'$' or $t$-channel leptoquark exchange can be parametrized 
as 4-fermion interactions if $\sqrt{s}<< \Lambda$. 

In the gauge sector 
the trilinear gauge boson vertex $\gamma WW$ can be sensitive to new 
physics via new particles included in the vertex loop corrections
\cite{Monig:2005ge}.   It has become the practice to parametrize 
the trilinear gauge boson vertices in terms of $k_\gamma$ and 
$\lambda_\gamma$.  
M\"onig and Sekaric presented results of a 
detailed simulation including polarization and backgrounds 
of $\gamma\gamma\to W^+W^-\to q\bar{q}q\bar{q}$ at 
$\sqrt{s}_{ee}=500$~GeV \cite{Monig:2005ge}.   
Their results on 
$k_\gamma$ and $\lambda_\gamma$ sensitivities comparing $e^+e^-$, 
$e\gamma$ and $\gamma\gamma$ modes for $\sqrt{s}_{ee}=500$~GeV
and ${\cal L}_{int}=1000$~fb$^{-1}$
are shown in Table \ref{monig}.

\begin{table}[htb]
\begin{center}
\caption{Comparison of the $\kappa_{\gamma}$ and
  $\lambda_{\gamma}$ sensitivities at $\gamma e$-, $\gamma\gamma$- and
  $e^{+}e^{-}$-colliders estimated at $\sqrt{s_{ee}}=500$ GeV using the
  polarised beams. In case of photon colliders, the background and the pileup
  are included. ($^*$) denotes the estimation at the generator
  level. } 
\label{monig}
\begin{tabular}{|c||c||c|c||c|} \hline
& \multicolumn{4}{|c|}{$\sqrt{s_{ee}}=500$ GeV}   \\ \hline\hline
LEFT & {$\gamma e$} & \multicolumn{2}{|c||}{$\gamma\gamma$} & $e^{+}e^{-}$  \\ \hline\hline
Mode & Real/Parasitic $|J_{Z}|=3/2$ & $|J_{Z}|=2$ & $J_{Z}=0$ & $|J_{Z}|=1$ \\ \hline
$\int{\cal L}\Delta t$ & 160 fb$^{-1}$/230 fb$^{-1}$ & \multicolumn{2}{|c||}{1000 fb$^{-1}$} & 500 fb$^{-1}$  \\ \hline
${{\Delta}L}$ & \multicolumn{1}{|c||}{0.1$\%$} & 0.1$\%$ & 1$\%$ & - \\ \hline\hline
${\Delta}{\kappa}_{\gamma}{\cdot}10^{-4}$ & 10.0/11.0 & 7.0 & 27.8 & 3.6$^*$  \\ \hline
${\Delta}{\lambda}_{\gamma}{\cdot}10^{-4}$ & 4.9/6.7 & 4.8 & 5.7 & 11.0$^*$ \\ \hline
\end{tabular}
\end{center}
\end{table}

\begin{figure*}[t]
\centering
\includegraphics[width=65mm]{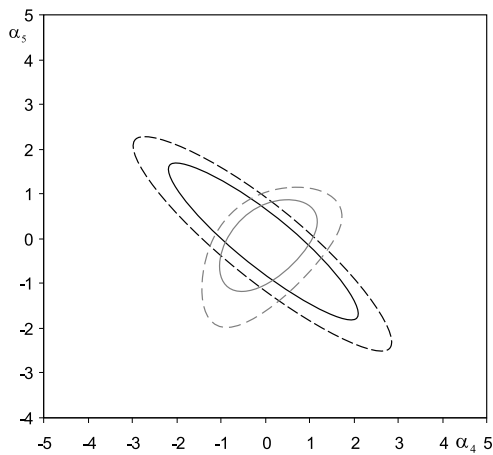}
\caption{Comparison of estimated sensitivities for $\alpha_4$ and 
$\alpha_5$ at
  $\sqrt{s}=1000$GeV from weak boson scattering (black) and triple boson
  production (gray). Lines represent $68\%$ (full) and $90\%$ (dashed)
  confidence level contours.
From Krstonosic, M\"onig, Beyer, Schmidt, and Schr\"oder
\cite{Krstonosic:2005qp}.} 
\label{krstonosic}
\end{figure*}

A strongly interacting weak sector would manifest itself 
in weak boson scattering, in particular in the quartic couplings.  In 
the chiral Lagrangian parametrization one operator of interest is 
(for other operators not shown see for example 
Ref.~\cite{Krstonosic:2005qp,Kilian:2005bz}): 
\begin{equation}
{\cal L}_4= {{\alpha_4}\over{16\pi^2}} Tr(V_\mu V_\nu)\; Tr(V^\mu V^\nu)
\end{equation}
Again, one can calculate the values of these coefficients for specific 
models so that the the values and patterns of the coefficients $\alpha_i$ 
will codify the underlying new physics.  Precision measurements of these 
coefficients will be needed to disentangle the underlying physics. 
The quartic vertices can be studied in numerous gauge boson scattering 
processes such as $e^+e^-\to \nu\bar{\nu}W^+W^-$ but also in triboson 
production processes such as $e^+e^-\to W^+W^-Z$.  An important goal 
is to produce a full and consistent set of limits using all possible 
processes.  Once this is done 
one can produce a strategy of measurements to constrain various models 
of new physics.  To this end Krstonosic \cite{Krstonosic:2005qp}
presented results of a new analysis of gauge boson scattering and 
triple gauge boson production. Their results are summarized in 
Fig.~\ref{krstonosic}.  To obtain these bounds on $\alpha_4$ and 
$\alpha_5$ they assumed the same integrated luminosity 
 and $80\%$ left $e^-$ and $40\%$ right $e^+$ polarization for scattering 
and $80\%$ right $e^+$ and $60\%$ left $e^-$ polarization for triple 
production.  The same luminosity based conclusion was made
 after comparison of $e^+e^-$ and $e^-e^-$ running modes leaving the
 experimental physicist several ways to achieve the desired precision.  

\section{CONCLUSIONS}

The ILC is an instrument for making precision measurements.  
These are needed to 
disentangle the underlying new physics that may be hinted at or 
unearthed elswhere.  For example, if an $s$-channel resonance were
discovered at the LHC, the ILC would be needed to make precision 
measurements of its properties.  Without these measurements, 
complementary to those of the LHC, it will be unlikely we will know 
the underlying theory.  Another example is the discovery of a light 
Higgs boson at the LHC.  Again, precision measurements at the ILC are 
needed to determine its origins.  Some recent examples that have been 
discussed in the literature and presented at this workshop are 
distinguishing between a SM Higgs, SUSY, ADD, etc. These examples take 
advantage of a prior discovery at the LHC to extract further 
information using precision measurements at the ILC.  There are other 
examples for which the ILC has a higher reach than the LHC via 
indirect effects such as interference of new interactions with the SM 
or via loop contributions to effective interactions.  

An important task for our community is to continue to strengthen the 
case that the ILC is needed, especially in the era of the LHC.  
To do this we shouldn't forget the LHC.   
Working on LHC physics is needed to understand its strengths and 
weaknesses thereby pointing out where the ILC can contribute crucial 
measurements.  The rewards of 
our efforts might be in a press release by the ILC Director 
proclaiming ``This result will send theorists back to their drawing 
boards'', and what could be more exciting than that!

\begin{acknowledgments}
The author thanks contributors to the BSM sessions for helpful 
conversations and communications and H. Logan and T. Rizzo for helpful 
comments.  This work was supported by 
the Natural Sciences and Engineering Research Council of Canada.
\end{acknowledgments}


\end{document}